# A Secure Communication Game with a Relay Helping the Eavesdropper

Melda Yuksel
EEE Department
TOBB University of Economics and Technology
Ankara, Turkey
yuksel@etu.edu.tr

Xi Liu
ECE Department
Polytechnic Institute of NYU
Brooklyn, NY 11201
xliu02@students.poly.edu

Elza Erkip
ECE Department
Polytechnic Institute of NYU
Brooklyn, NY 11201
elza@poly.edu

*Abstract*—In this work a four terminal Gaussian network composed of a source, a destination, an eavesdropper and a jammer relay is studied. The jammer relay does not hear the source transmission. It assists the eavesdropper and aims to decrease the achievable secrecy rates. The source, on the other hand, aims to increase the achievable secrecy rates. Assuming Gaussian strategies at the source and the jammer relay, this problem is formulated as a two-player zero-sum continuous game, where the payoff is the achieved secrecy rate. For this game the Nash Equilibrium is generally achieved with mixed strategies. The optimal cumulative distribution functions for the source and the jammer relay that achieve the value of the game, which is the equilibrium secrecy rate, are found.

## I. INTRODUCTION

In wireless communications, messages are broadcasted, and any transmission can be overheard by nearby nodes. If illegitimate, passive listeners trying to *understand* messages, known as eavesdroppers, are present in the environment, then all confidential information such as user IDs, or passwords, become vulnerable and can be identified. Therefore, security against eavesdropping is an essential system requirement for all wireless communication applications.

Security against eavesdropping using information theoretic principles was first considered in [1]. In [2], the wire-tap channel was studied for the degraded case, when the eavesdropper's received signal is a degraded version of the legitimate receiver's observation. This model was extended to less noisy and more capable wire-tap channels in [3]. The Gaussian case was studied in [4]. Secure wireless networking applications motivated recent interest in multi-user secret communication. In [5] multiple access channels with confidential messages were studied. The Gaussian multiple-access and two-way wire-tap channels were investigated in [6] and [7].

Secrecy in relay channels was studied in [8], [9] and [10]. In these networks, the messages are to be kept secret from the relay node itself. On the other hand, the relay channel with an external eavesdropper is investigated in [11], [12] and [13]. In [11], the authors suggest the noise forwarding scheme, where the relay transmits dummy codewords that can be decoded at the destination. While sending dummy codewords does not hurt the legitimate communication, it increases the confusion at the eavesdropper, and hence helps achieve a higher secrecy rate. Noise forwarding scheme is similar to cooperative jamming [7], in which one of the users in the system injects noise to increase achievable secrecy rates in multi-access and two-way channels. Finally, the paper [14] ties [11] and [7] together, and shows that the helper can choose between sending structured codewords and pure noise to increase achievable secrecy rates even further.

Unlike these works, in which user's/relay's transmissions aim to help the legitimate source-destination communication, the relay can aim to help the eavesdropper. In a general multiple terminal network this is a possible scenario, if a relay is captured by an adversary [15]. In this case, the relay aims to decrease the secrecy rates. In [16] we studied this problem for orthogonal transmission between the source and the relay and compared secrecy rates achievable with amplify-and-forward, decode-and-forward and compress-and-forward protocols with the direct transmission.

In this work we revisit the problem of achieving secrecy when the relay helps the eavesdropper. We assume that the relay does not hear the source transmission and can therefore transmit simultaneously with the source. This assumption makes noise forwarding/jamming type protocols meaningful, unlike the orthogonal communication scenario considered in [16]. However, as the relay does not hear the source transmission, correlated jamming as in [17] or [18] is not possible. As the *jammer* relay helps the eavesdropper, it tries to minimize the secrecy rate, while the source transmits to maximize the secrecy rate. We formulate this problem as a two-player zero-sum continuous game. We find the Nash equilibrium of this game, which in general is obtained by a mixed strategy.

In the next section we state the problem. In Section III we describe the game theoretic formulation. In Section IV we present the Nash equilibrium solution of the game, and in Section V we conclude.

## II. PROBLEM STATEMENT

We investigate the four terminal network composed of a source, a destination, an eavesdropper and a jammer relay denoted by S, D, E and R respectively. The network under investigation is shown in Fig. 1.

[1]This material is based upon work partially supported by NSF Grant No. 0635177, by the Center for Advanced Technology in Telecommunications (CATT) of Polytechnic Institute of NYU.

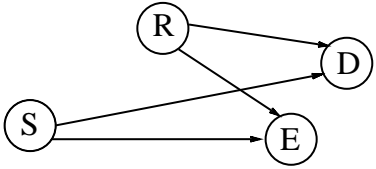

Fig. 1. The system model shows the source (S), the destination (D), the eavesdropper (E) and the jammer relay (R). The jammer relay aims to assist the eavesdropper.

The received signals at the destination and the eavesdropper are

$$Y_{D,i} = h_{SD}X_{S,i} + h_{RD}X_{R,i} + Z_{D,i}$$
$$Y_{E,i} = h_{SE}X_{S,i} + h_{RE}X_{R,i} + Z_{E,i},$$

where $X_{S,i}$ and $X_{R,i}$ are the signals the source and the relay transmit at time $i$, $i = 0, ..., n$. The complex additive Gaussian noise at the destination and at the eavesdropper are respectively denoted as $Z_{D,i}$ and $Z_{E,i}$ and are independent and identically distributed (i.i.d.) with zero mean and unit variance. The channel gains between node $k$ and node $l$ are shown as $h_{kl}$, $k = S, R$, $l = D, E$. All channel gains are fixed and assumed to be known at all nodes. The source and the jammer relay have average power constraints

$$\frac{1}{n}\sum_{i=1}^{n} E[X_{S,i}^2] \leq P_S, \quad \frac{1}{n}\sum_{i=1}^{n} E[X_{R,i}^2] \leq P_R.$$

For convenience we will write $\gamma_{kl} = |h_{kl}|^2 P_k$, $k = S, R$, $l = D, E$, to indicate the received power at node $l$ due to node $k$.

The source aims to send the message $W$ securely to the destination in $n$ channel uses. The equivocation rate is defined as [3] the entropy of the message $W$ given the eavesdropper's observation $Y_{E,1}^n$, $H(W|Y_{E,1}^n)/n$. Note that the equivocation rate is equal to $H(W)/n$, if the eavesdropper's observation is independent from the message $W$. On the other hand, if the observation $Y_{E,1}^n$ is enough to recover the message $W$ with arbitrary small probability of error as $n$ approaches infinity, then the equivocation rate is zero. The *perfect secrecy* rate, $R_s$ is defined as the maximum information rate such that the equivocation rate is equal to the transmission rate; i.e. $\lim_{n\to\infty} H(W|Y_{E,1}^n)/n = \lim_{n\to\infty} H(W)/n$, and the probability of error at the destination $P_e^{(n)}$ approaches zero, as $n$ approaches infinity.

In this problem the source and the jammer relay have opposing interests. The former wants to increase the secrecy rate, and the latter wants the decrease $R_s$. Thus, this problem constitutes a zero-sum game, where the utility is the perfect secrecy rate. The source and the jammer relay make their decisions simultaneously, and hence the game is strategic. The source chooses a strategy $\xi$ and the jammer relay chooses a strategy $\eta$ simultaneously. The secrecy rate, or the payoff, is a function of both $\xi$ and $\eta$. If a certain positive secrecy rate, $R_s(\xi, \eta)$ is achieved, then the source node's payoff is equal to $R_s(\xi, \eta)$ and the relay's payoff is equal to $-R_s(\xi, \eta)$.

In the next section, we define the strategy spaces for $\xi$ and $\eta$ and calculate the payoff function $R_s(\xi, \eta)$.

## III. GAME THEORETIC FORMULATION

For the communication to take place the source node maps its uniformly chosen message $W$ to a channel codeword $X_{S,1}^n$. The rate of this communication will be specified later.

If the jammer relay were a jammer only, then it would simply transmit unstructured noise. However, in this problem it is not merely a jammer, but aims help the eavesdropper to ensure no secret communication takes place. Then, we consider two options for the jammer relay: it can either send unstructured noise or structured codewords. Unstructured noise is useful as it harms the legitimate communication, yet it also harms the eavesdropper. Structured codewords have the potential to help the eavesdropper more.

As the jammer relay does not hear the source node, structured codewords can only carry dummy information. However, structured codewords still force the source to transmit at lower rates to ensure decodability at the destination and can affect the secrecy rates. Thus, we assume that the jammer relay generates dummy codewords and sends the channel codeword $X_{R,1}^m$ or simply forwards noise. In [19], it is argued that there is no loss of generality in assuming same codeword lengths for both players in an information theoretic game on interference channels. Even though in our problem the players are adverseries, we assume $m = n$. We assume all nodes in the system know both the source and the jammer relay's strategies, including the codebooks.

When both the source and the jammer relay send structured codewords, the problem becomes similar to a multiple-access channel with an external eavesdropper. In a multiple-access channel, both the transmitters need to be decoded at the destination. However, in this problem, the jammer relay only sends dummy codewords, and does not need to be decoded either at the destination or at the eavesdropper.

Observe that when the jammer relay sends Gaussian codewords/noise with full power, the best distribution for the source is the Gaussian distribution [4] with zero mean and variance $P_S$. On the other hand, when the source sends Gaussian codewords with zero mean and variance $P_S$, then the jammer relay distribution that decreases the perfect secrecy the most is the Gaussian distribution, zero mean and variance $P_R$. Motivated by these observations, we assume that the source and the jammer relay choose their codebooks independent and identically distributed Gaussian with zero mean and variances $P_S$ and $P_R$ respectively. Then, the source strategy, denoted by $\xi$, is to choose the rate of the information $W$, and the jammer relay strategy, $\eta$, is to choose the rate of its dummy information. We argue below that structured codewords at the jammer relay also include the possibility of sending pure noise.

Under these assumptions, the destination can decode both the source and the jammer relay codewords if the rate pair $(\xi, \eta)$ is in $\mathcal{R}_{\text{MAC}}^{[D]}$

$$\mathcal{R}_{\text{MAC}}^{[D]} = \left\{ (\xi, \eta) \; \middle| \; \begin{array}{rcl} \xi & \leq & \log(1 + \gamma_{SD}) \\ \eta & \leq & \log(1 + \gamma_{RD}) \\ \xi + \eta & \leq & \log(1 + \gamma_{SD} + \gamma_{RD}) \end{array} \right\}. \quad (1)$$

However, the jammer relay only sends dummy codewords, and does not need to be decoded either at the destination or at the eavesdropper. If the destination cannot decode the relay codeword, it can simply treat it as noise. Thus, all $\xi$ rates in $\mathcal{R}_N^{[D]}$

$$\mathcal{R}_N^{[D]} = \left\{ \xi \, \Big| \, \xi \leq \log\left(1 + \frac{\gamma_{SD}}{1+\gamma_{RD}}\right) \right\} \quad (2)$$

are achievable as well. Overall, we say that the destination can decode $W$ with arbitrarily small probability of error, if $(\xi, \eta) \in \mathcal{R}^{[D]} = \mathcal{R}_{\text{MAC}}^{[D]} \bigcup \mathcal{R}_N^{[D]}$. Note that after taking the union, the individual constraint on $\eta$ in (1) is not needed anymore. We also define three other regions $\mathcal{R}_{\text{MAC}}^{[E]}$, $\mathcal{R}_N^{[E]}$, and $\mathcal{R}^{[E]}$ as in (1) and (2) and $\mathcal{R}^{[D]}$, but replacing all $D$ with $E$. Then for a fixed source and jammer relay rate pair $(\xi, \eta)$, the payoff function, $R_s(\xi, \eta)$, is equal to

$$R_s(\xi, \eta) = \begin{cases} 0, & \text{if } \begin{pmatrix} (\xi, \eta) \in \mathcal{R}^{[E]} \text{ or} \\ (\xi, \eta) \notin \mathcal{R}^{[D]} \end{pmatrix} \\ \max_{R_{S,d}} (\xi - R_{S,d}), & \text{if } \begin{pmatrix} (\xi, \eta) \in \mathcal{R}^{[D]} \text{ and} \\ (R_{S,d}, \eta) \notin \mathcal{R}^{[E]} \text{ and} \\ R_s(\xi, \eta) + R_{S,d} = \xi \end{pmatrix} \end{cases} \quad (3)$$

The proof of how this secrecy rate would be achieved is similar to [14], [20] and is skipped here.

An example is shown in Fig. 2 for the boundary regions $\mathcal{R}^{[D]}$ and $\mathcal{R}^{[E]}$ with the corner points defined as

$$(\Delta_S, \Delta_R) = \left(\log(1 + \gamma_{SD}), \log\left(1 + \frac{\gamma_{RD}}{1+\gamma_{SD}}\right)\right),$$

$$(\Omega_S, \Omega_R) = \left(\log\left(1 + \frac{\gamma_{SD}}{1+\gamma_{RD}}\right), \log(1+\gamma_{RD})\right),$$

$$(\delta_S, \delta_R) = \left(\log(1 + \gamma_{SE}), \log\left(1 + \frac{\gamma_{RE}}{1+\gamma_{SE}}\right)\right),$$

$$(\omega_S, \omega_R) = \left(\log\left(1 + \frac{\gamma_{SE}}{1+\gamma_{RE}}\right), \log(1+\gamma_{RE})\right).$$

For a fixed $(\xi, \eta)$, the secrecy rate defined in (3) corresponds to the horizontal distance between the point $(\xi, \eta)$ and the dashed line in Fig. 2, if $(\xi, \eta)$ is in between the solid and dashed lines. If $(\xi, \eta) \in \mathcal{R}^{[E]}$, that is inside the dashed line, then both the destination and the eavesdropper can reliably decode $W$, and the secrecy rate is zero. If $(\xi, \eta) \notin \mathcal{R}^{[D]}$, outside the solid line, the destination cannot decode the source message reliably. The secrecy rate is zero, because there is no reliable communication between the source and the destination. Because of this immediate drop in secrecy rates beyond the boundary of $\mathcal{R}^{[D]}$, the payoff function is discontinuous.

Note that the corner point $(\omega_s, \omega_R)$ is equivalent to sending unstructured Gaussian noise at the jammer relay. Thus, sending unstructured Gaussian noise is also covered in our model, although we arrived at the $R_s(\xi, \eta)$ function assuming structured codewords for the jammer relay.

Depending on $\gamma_{kl}$, the positions of the corner points with respect to each other change, and multiple cases arise. In this paper we investigate the case, where the conditions (i) $\log(1 + \gamma_{SE} + \gamma_{RE}) \leq \log(1 + \gamma_{SD} + \gamma_{RD})$, (ii) $\delta_S \leq \Delta_S$, (iii) $\Delta_R \leq \delta_R$, (iv) $\omega_S \leq \Omega_S \leq \delta_S$, and (v) $\delta_R \leq \Omega_R \leq \omega_R$ are all satisfied. The case shown in Fig. 2 satisfies all these conditions. Other cases require similar techniques, but have different solutions. Due to limited space, we do not explain those here.

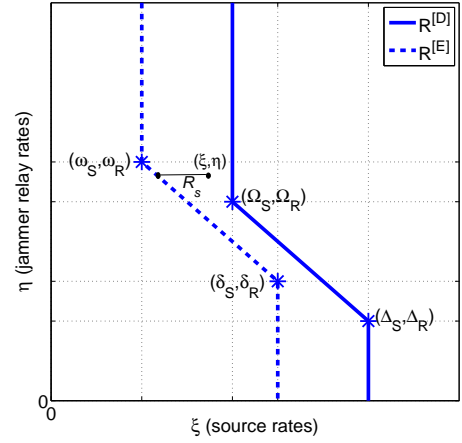

Fig. 2. Boundaries for $\mathcal{R}^{[D]}$ and $\mathcal{R}^{[E]}$, under conditions (i)-(v).

## IV. PLAYING THE GAME

In this section we explain pure and mixed strategy solutions to the zero-sum game and show that this problem has a mixed strategy solution. We also state the cumulative distribution functions for the source and jammer relay strategies and find the equilibrium secrecy rate, or the value of the game.

A zero-sum game has a pure strategy solution if [21]

$$\max_\xi \min_\eta R_s(\xi, \eta) = \min_\eta \max_\xi R_s(\xi, \eta).$$

However, if there is no $(\xi, \eta)$ that satisfies this equation, then no pure strategy Nash equilibrium exists, and a mixed strategy solution is needed.

*Lemma 1:* When $\gamma_{kl}$, $k = S, R$, $l = D, T$ satisfy the conditions (i)-(v), the two-player zero-sum game does not have a pure strategy solution.

*Proof:* In this game $\max_\xi \min_\eta R_s(\xi, \eta) = 0$, whereas $\min_\eta \max_\xi R_s(\xi, \eta) = \log(1 + \gamma_{SD} + \gamma_{RD}) - \log(1 + \gamma_{SE} + \gamma_{RE})$. These two values are not the same, hence a pure strategy solution does not exist. ∎

*Lemma 2:* The game defined in Lemma 1, is equivalent to the game played over the square, where the source and jammer relay strategies are respectively restricted to the compact intervals $\xi \in [\Omega_S, \Omega_S + L]$ and $\eta \in [\delta_R, \delta_R + L]$, where $L$ is the edge length $L = \Omega_R - \delta_R$.

*Proof:* To prove this we eliminate comparable and inferior strategies for the source and the jammer relay. First note that

$$R_s(\xi, \eta) \leq R_s(\Omega_S, \eta), \text{ for } \xi < \Omega_S, \text{ and } 0 < \eta.$$

In other words, as both players are rational, the source never chooses rates $\xi < \Omega_S$. Similarly, the source node never chooses its rate larger than $\Delta_S$, as the secrecy rate $R_s(\xi, \eta) = 0$ no matter what the relay action is. On the other hand,

$$R_s(\xi, \eta) \geq R_s(\xi, \Omega_R), \text{ for } \xi > 0, \text{ and } \eta > \Omega_R.$$

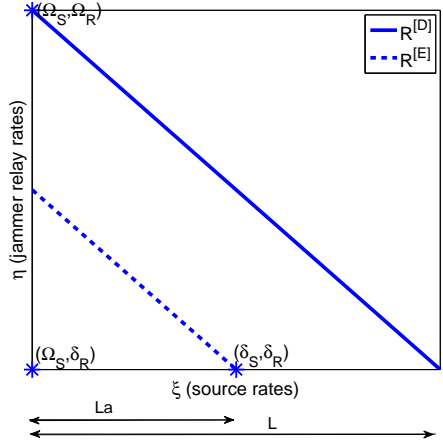

Fig. 3. The equivalent game for the game in Fig. 2.

Thus, for the jammer relay choosing any rate larger than $\Omega_R$ is equivalent to choosing rate equal to $\Omega_R$. The strategies $\eta > \Omega_R$ are comparable to the strategy $\eta = \Omega_R$, and we can omit the strategies $\eta > \Omega_R$. Similarly,

$$R_s(\xi, \eta) \geq R_s(\xi, \delta_R), \text{ for } \xi > 0, \text{ and } \eta < \delta_R.$$

Thus, the strategies $\eta < \delta_R$ are inferior to $\eta = \delta_R$ and the jammer relay never chooses its rate less than $\delta_R$. Finally, in this reduced game, the source node does not choose its rate larger than $\log(1 + \gamma_{SD} + \gamma_{RD}) - \delta_R$, as this choice makes its payoff equal to zero. In other words,

$$R_s(\xi, \eta) \leq R_s(\log(1 + \gamma_{SD} + \gamma_{RD}) - \delta_R, \eta),$$

for $\xi > \log(1 + \gamma_{SD} + \gamma_{RD}) - \delta_R$ and $\delta_R < \eta < \Omega_R$. These strategy eliminations result in the desired reduced game. ∎

The reduced game based on Lemma 2 is shown in Fig. 3. We next describe how to solve for the value of this reduced game. The parameter $a$ is defined as $a = (\delta_S - \Omega_S)/L$.

*Theorem 1:* Suppose $a \in [k/(k+1), (k+1)/(k+2)]$, for some integer $k \geq 0$. Then the reduced game in Lemma 2 has the Nash equilibrium secrecy rate $R_s^* = L\alpha(1-a)$, where $\alpha = g_k(a)$, and is achieved with cumulative distribution functions for the source and the jammer relay $F_\xi(\xi)$ and $F_\eta(\eta)$, respectively. The c.d.f.s $F_\xi(\xi)$, $F_\eta(\eta)$ and the function $g_k(a)$ can be analytically computed for a given $k$.

*Proof:* First we normalize the edge lengths of the square in Fig. 3 and use the game-theoretic techniques in [21] to solve the continuous game played over the unit square. As the space is limited, the proof is omitted here and will be presented in the journal submission. ∎

Since there are infinitely many intervals for $a$ in Theorem 1 (corresponding to each nonnegative integer $k$) we present some examples here. When $0 \leq a \leq 1/2$, $\alpha = g_0(a) = \frac{e^{-1/(1-a)}}{1 - \frac{a}{1-a}e^{-1}}$ and the c.d.f. for the source node is

$$F_\xi(\xi) = \begin{cases} \alpha e^{\frac{\xi - \Omega_S}{L(1-a)}}, & \text{if } \Omega_S \leq \xi \leq \Omega_S + L(1-a) \\ \alpha \left[(1+e^{-1})e^{\frac{\xi-\Omega_S}{L(1-a)}} - \frac{1}{1-a}(\frac{\xi-\Omega_S}{L})e^{\frac{\xi-\Omega_S}{L(1-a)}-1}\right], \\ \quad \text{if } \Omega_S + L(1-a) \leq \xi \leq \Omega_S + L \end{cases}$$

The c.d.f. for the jammer relay is the same as $F_\xi(\xi)$ if $\xi$ and $\Omega_S$ are replaced with $\eta$ and $\delta_R$ respectively. When $1/2 < a \leq 2/3$,

$$\alpha = g_1(a) = \frac{e^{-1/(1-a)}}{1 + e^{-1} + 2e^{-2} - \frac{e^{-1}+2e^{-2}}{1-a} + \frac{e^{-2}}{2(1-a)^2}},$$

and the c.d.f. for the source node is given in (4). The c.d.f. for the jammer relay is the same as (4) if $\xi$ and $\Omega_S$ are replaced with $\eta$ and $\delta_R$ respectively.

Although, it is always possible to obtain the optimal c.d.f.s and $\alpha$ analytically for a given $a$, a general closed form expression does not exist. Therefore, it is important to find a practical way to calculate the value of the game.

*Theorem 2:* Suppose we form a discrete approximation of the reduced game in Lemma 2, by taking $(T+1)^2$ samples; obtained by dividing the square into a uniform grid. Then the discrete source strategies are $\xi_l = \omega_S + Ll/T$, the relay strategies are $\eta_l = \delta_R + Ll/T$, and the payoff function is $R_s(l) = R_s(\xi_l, \eta_l)$, where $l = 0, 1, ..., T$. The value of this discrete game can be obtained using linear programming. Furthermore, for a chosen T, difference between values of the discrete and the continuous game is at most $2\sqrt{(2)}L/T$.

*Proof:* We use the approximation techniques suggested in [22]. Due to space limited space, the proof is omitted. ∎

### A. Achievability

If the solution of the game is a pure strategy, the achievability follows using the arguments in Section III. As the solution is mixed, it is also important to explain how information theoretically the Nash equilibrium is attained.

In a mixed strategy the players randomize their actions over a set of strategies with a certain probability distribution. The players act repeatedly and ignore any strategic link that may exist between plays. They also know each other's probability distribution functions, and hence formulate their own actions. In the game defined in this paper, when a mixed strategy solution is needed, the source node assumes a variable rate

$$F_\xi(\xi) = \begin{cases} \alpha e^{\frac{\xi-\Omega_S}{L(1-a)}}, & \Omega_S \leq \xi \leq \Omega_S + L(1-a) \\ \alpha \left[(1+e^{-1})e^{\frac{\xi-\Omega_S}{L(1-a)}} - \frac{1}{1-a}(\frac{\xi-\Omega_S}{L})e^{\frac{\xi-\Omega_S}{L(1-a)}-1}\right], & \Omega_S + L(1-a) \leq \xi \leq \Omega_S + 2L(1-a) \\ \alpha e^{\frac{\xi-\Omega_S}{L(1-a)}}\left[1 + e^{-1} + 2e^{-2} - \frac{1}{1-a}(e^{-1}+2e^{-2})(\frac{\xi-\Omega_S}{L}) + \frac{(\frac{\xi-\Omega_S}{L})^2 e^{-2}}{2(1-a)^2}\right], & \Omega_S + 2L(1-a) \leq \xi \leq \Omega_S + L \end{cases} \quad (4)$$

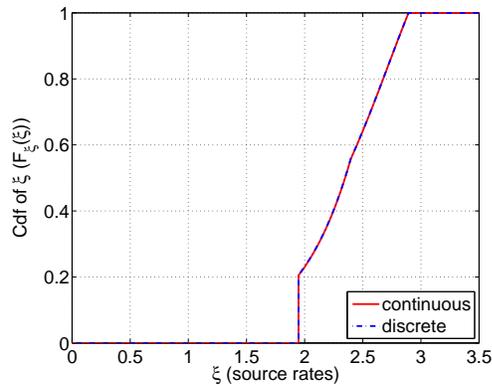

Fig. 4. Optimal c.d.f. for the source, for the continuous and discrete games when $a = 0.5255$

scheme, similar to the one adopted for fading eavesdropper channels [23].

In this variable rate scheme, the source generates a total of $2^{nBE(\xi)}$ codewords, where $B$ is the number of blocks the game is played, and $E(\xi)$ is the expected rate for the source node, expectation calculated over the joint c.d.f. $F_\xi(\xi)F_\eta(\eta)$. The source uses these codewords to form a secure code that conveys $nBR_s^*$ bits of information in $B$ blocks [2], where $R_s^*$ is the value of the game or the Nash secrecy rate. In each block, the source randomly chooses a rate $\xi$ according to $F_\xi(\xi)$ and transmits $n\xi$ bits of the codeword chosen to represent the secure information. Similarly, the jammer relay chooses a rate $\eta$ according to $F_\eta(\eta)$. Since the eavesdropper cannot improve its mutual information more than $\xi$, as in the variable rate case of [23], (3) is still valid and $R_s^*$ is attained as both $n$ and $B$ approach infinity.

### B. Examples

Now, we compare the optimal strategies for the discrete and continuous games using the solutions in Theorems 1 and 2, when $1/2 < a \leq 1$. We assume $|h_{SD}| = 1$, $|h_{SR}| = 1/3$, $|h_{RD}| = 1/2$, and $|h_{SE}| = |h_{RE}| = 2/3$. The source and the jammer relay power constraints are $P_s = P_r = 10$. Then $a = 0.5255$, $L = 0.946$ and $\alpha = 0.20484$. We find the equilibrium secrecy rate of the continuous game as 0.092 bits/channel use. To use the discrete approximation we set $T = 400$. Choosing this sample size, we analytically prove that the difference between the value of discrete and continuous games is at most 0.007. Yet, the actual difference is much smaller and we find that the value of the discrete game as 0.0923 bits/channel use. Note that these values are much smaller than the no jammer relay case, for which the secrecy rate is equal to 1.0146 bits/channel use. The optimal c.d.f. for the source for the continuous and discrete games are very close to each other and are shown in Fig. 4. We observe that $F_\xi(\xi)$ is zero if $\xi < 1.947 = \Omega_S$, and is 1 if $\xi > 2.893 = \Omega_S + L$. Note that in the reduced game, sending Gaussian noise with full power is still one of jammer relay's possible strategies.

## V. CONCLUSION

In this paper we investigated the four terminal network with a source, a destination, an eavesdropper and a jammer relay. The source and the jammer relay have conflicting interests. The former aims higher secrecy rates, whereas the latter aims lower secrecy rates. Due to this conflict, we formulate this problem as a two-player zero-sum continuous game, and find the optimal solution for the source and the jammer relay to be mixed strategies. We find the equilibrium secrecy rate of the game, in addition to optimal cumulative distribution functions for the source and the jammer relay. We also find a discrete approximation to the continuous game, whose value can be made arbitrarily close to the value of the continuous game. Our results show that the presence of the jammer relay decreases the secrecy rates significantly. As for future work, the cases in which the jammer relay hears the source transmission remain open.